\documentclass[12pt, preprint]{aastex}
\slugcomment{Submitted to ApJ}
\shorttitle{The VLA-COSMOS Survey}
\shortauthors{Bondi et al.}

\begin{document}

\title{The VLA-COSMOS Survey: \\
III. Further Catalog Analysis and the Radio Source Counts}

\author{M. Bondi\altaffilmark{1}, P. Ciliegi\altaffilmark{2}, 
E. Schinnerer\altaffilmark{3}, V. Smol\v{c}i\'{c}\altaffilmark{3},
K. Jahnke\altaffilmark{3}, C. Carilli\altaffilmark{4},
G. Zamorani\altaffilmark{2}}

\altaffiltext{1}{INAF - Istituto di Radioastronomia, 
Via Gobetti 101, I-40129 Bologna, Italy}
\altaffiltext{2}{INAF - Osservatorio Astronomico di Bologna, Via Ranzani 1,
I-40127, Bologna, Italy}
\altaffiltext{3}{Max-Planck-Institut f\"ur Astronomie, K\"onigsthul 17, D-69117 
Heidelberg, Germany}
\altaffiltext{4}{National Radio Astronomy Observatory, P.O. Box O, Socorro, 
NM 87801-0387, U.S.A.}

\begin{abstract}
The VLA-COSMOS Large Project has imaged the 2 deg$^2$ COSMOS field with a
resolution of 1.5 arcsec and a sensitivity of about 11 $\mu$Jy (1$\sigma$)
yielding to a catalog of $\sim 3600$ radio sources. In this paper
 we present a further analysis of the VLA-COSMOS Large Project catalog 
of radio sources
aimed to: 1) quantify and  correct for the effect of bandwidth smearing in the 
catalog, 2) determine the incompleteness produced by
the noise bias and the resolution bias in the new catalog and
3) derive the radio source counts at 1.4 GHz.

The effect of bandwidth smearing on the radio sources in the catalog 
was quantified
comparing the peak and total flux densities in the final mosaic and in
each of the individual pointings where the source was closest to the center
of the field.
We find that the peak flux densities in the original VLA-COSMOS Large Project
catalog have to be divided by a factor about 0.8 or 0.9, depending on the 
distance from the mosaic center.

The completeness of the radio catalog has been
tested using samples of simulated radio sources with different angular
size distributions.
These simulated sources have been added to the radio image
and recovered using the same techniques used to  produce the radio 
catalog. The fraction of missed sources as a function of the total flux
density is a direct measure of the  incompleteness.

Finally, we derived the radio source counts down to  60 $\mu$Jy
with unprecedented good statistics. Comparison to the findings of
other surveys shows good agreement in the flux density range 0.06-1 mJy
confirming the upturn at $\sim 0.5$ mJy and a possible decline of the
source counts below $\sim 0.1$ mJy.

\end{abstract}

\section{Introduction}

In recent years, we have experienced a renaissance of deep radio surveys,
usually associated with multi-waveband observational programmes mainly 
designed to 
probe the formation and evolution of galaxies and Super Massive Black Holes 
(SMBH).
These surveys, mostly carried out at 1.4 GHz, have enabled the study of 
sub-mJy and $\mu$Jy radio source
populations (hereafter $\mu$Jy population) and in particular 
the determination, with robust statistics, of the radio source counts down to 
a few tens 
of $\mu$Jy \citep[e.g.][]{Hopk98,Rich00,Bond03,Cili03,Hopk03,Seym04,Huyn05,
Pran06,Foma06,Simp06,Ivis07,Bond07}.

While it is clear that the $\mu$Jy population is a mixture of different
classes of objects (radio loud and radio quiet AGN, starburst galaxies,
spirals), the exact contribution of these different classes to 
the radio source counts is still not very well 
established and often dependent on the optical limit of the spectroscopic
follow-up \citep{Benn93,Hamm95,Geor99,Grup99,Pran01,Afon05,Foma06,Simp06}.
The role played by the cosmological evolution of the different classes of 
objects is even more uncertain.
Recently, \citet{Smol06} have developed a new classification method to
separate star forming  galaxies and AGN based on 
photometric rest-frame colors for local galaxy samples. 
The application of this classification 
method to the VLA-COSMOS sources \citep{Smol08a} shows that the radio 
population above $\sim$ 50
$\mu$Jy is a mixture of roughly 30--40\% of star forming galaxies
and 50--60\% of AGN galaxies, with a minor contribution ($\sim$ 10\%) of QSOs.

The international COSMOS (Cosmic Evolution) survey \citep{Scov07a}
is designed to probe the correlated evolution of galaxies, star
formation, AGN up to high redshift. The survey includes
state of the art imaging data {covering the entire wavelength range} from the 
X-rays to the radio domain 
\citep{Hasi07,Zamo07,Tani07,Scov07b,Capa07,Sand07,Bert07,Schi04,Schi07} 
supported by extensive optical 
spectroscopic campaigns using  mainly the VLT/VIMOS and the 
Magellan/IMACS 
instruments \citep{Lill07,Trum07}.

The radio observations of the COSMOS field were done with the VLA in A
and C configuration and are  described in Schinnerer et al.
(2004, 2007). The whole 2 deg$^2$ field was imaged  
in mosaic mode with a resolution of 1.5 arcsec. 
The sensitivity is uniform in the inner
1 deg$^2$ with an average rms value of 10.5 $\mu$Jy/beam and it increases
almost regularly to the outer regions of the 2 deg$^2$ field.
From the mosaic a catalog of radio sources, the VLA-COSMOS Large Project
Catalog, was
extracted. The catalog lists $\sim 3600$ radio sources selected 
above the local $4.5\sigma$ threshold over an area of 2 deg$^2$. 
A full description of the observations, data reduction and catalog
extraction is given in Schinnerer et al. (2007, hereafter S07).

In this paper we present a further analysis on the VLA-COSMOS Large Project
Catalog
aimed to quantify the impact of bandwidth smearing (Section 2) and the survey 
completeness (Section 3). Finally, we derive the radio source counts and
compare them with those derived from similar surveys (Section 4).

A more detailed interpretation of the radio data is left to
other papers which are currently in preparation. In particular, the optical 
identifications of the VLA-COSMOS radio sources are
presented by \citet{Cili08}, the fraction of star-forming galaxies and AGN
in the sub-mJy population of the VLA-COSMOS field is discussed in 
\citet{Smol08a} and the dust un-biased cosmic star formation history 
is derived from the 1.4 GHz radio data in \citet{Smol08b}.

\section{Bandwidth smearing correction}
Bandwidth smearing (or chromatic aberration) affects all synthesis 
observations made with a finite width of the receiver channels. 
Imaging sources at large distances from the phase center can yield to radial 
smearing, increasing the source size and reducing the peak flux density,
while the integrated flux density is conserved. The image smearing is 
proportional to the fractional bandwidth (the bandwidth divided by the central 
frequency of observation) and to the distance from the phase center in units 
of the synthesized beam \citep[e.g.][]{Thom99}.
For single pointing observation the bandwidth 
smearing simply increases with the radial distance from the 
phase centre. For a mosaic image, produced by multi-pointing observations, 
the contribution from all the pointings has to be taken into account and the 
resulting radial pattern will be much more complicated depending also
on the adopted spacing pattern of the  individual pointings.

\begin{figure}
\plotone{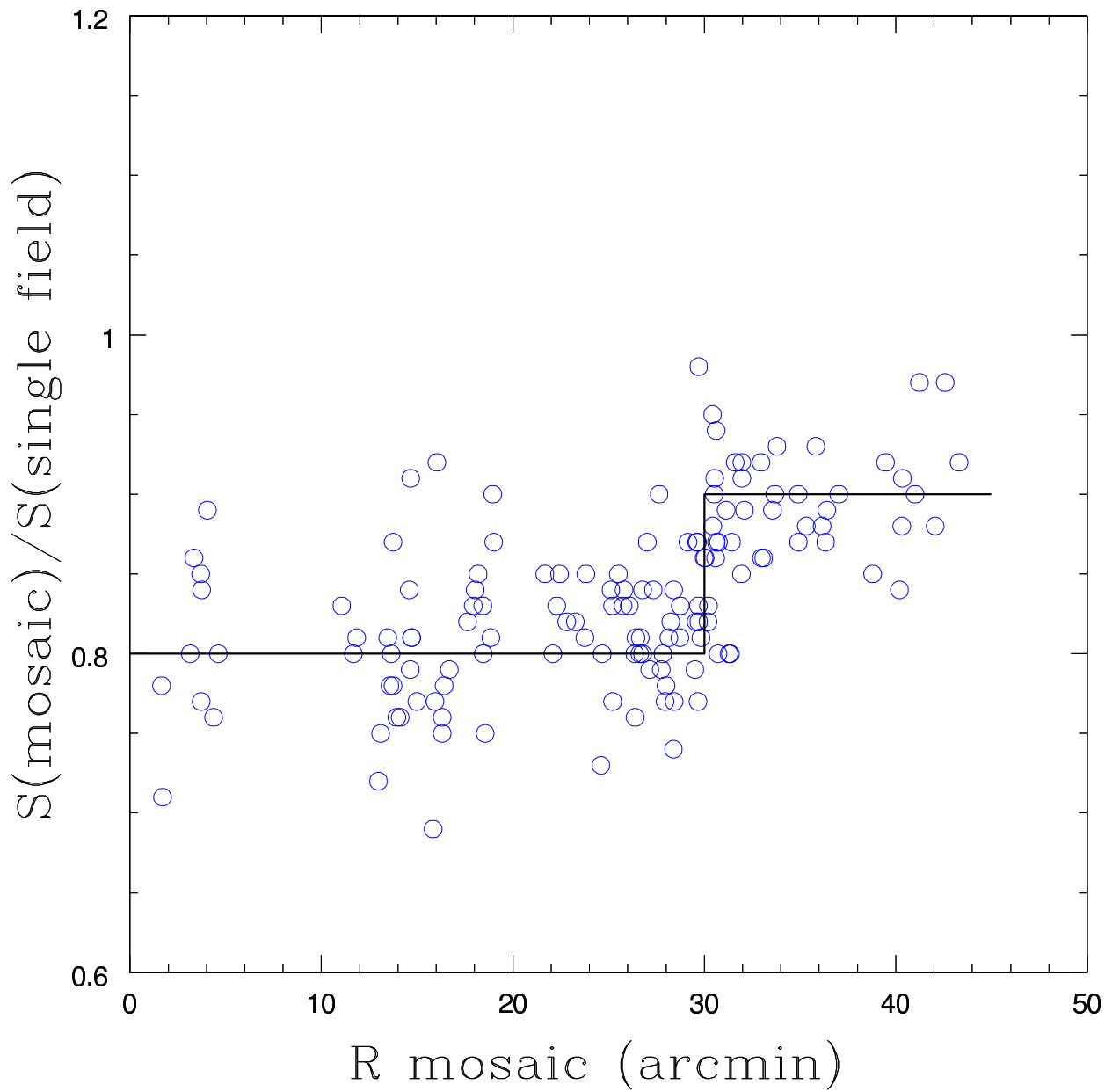}
\caption[]
{Ratio between the peak flux densities in the 
final mosaic and in the individual pointing where the source is within
$5\arcmin$ from the center versus
the radial distance from the center in the final mosaic.
Only sources with peak flux density greater than 0.2 mJy/beam and sizes
smaller than 4$\arcsec$ are plotted. The straight lines at 
S(mosaic)/S(single pointing)= 0.8 and 0.9 for R $<30\arcmin$ and 
R $>30\arcmin$ show the adopted bandwidth smearing correction factors.}
\label{bw_fig1}
\end{figure}

\placefigure{bw_fig1}

In order to quantify the effect of bandwidth smearing on the VLA-COSMOS 
observations we ran the same procedure that produced the radio catalog on
each of the 23 individual pointings (see S07).
For the strongest sources we compared their peak and total flux densities in
the final mosaic with the corresponding peak and total flux densities in the
pointing where the sources are closest to the field center.
The total flux density of each source in the mosaic and in the individual
pointings are in very good agreement (the median value of the ratio is 1.03 
with an r.m.s. dispersion of 0.04)
confirming that the total flux density is properly recovered.
On the other hand, 
as expected for observations affected by bandwidth smearing, the peak flux
densities are underestimated in the final mosaic with respect to the peak
fluxes
in the individual pointings where the sources are closest to the field center. 
To quantify this effect
we selected only sources which  are within a
radius $R_{\rm min}\le 5\arcmin$ from the field center in, at least, one
individual pointing. The peak flux density of a source  
within $5\arcmin$
from the field center is almost unaffected (less than 5\% for the
VLA-COSMOS observations) 
by bandwidth smearing and can be compared
with the peak flux in the mosaic. 
The image of the source in the mosaic is
the combination of all the images in the individual pointings where the
source is within the cut-off radius ($16.8\arcmin$, see S07).

In Figure~\ref{bw_fig1} we show the peak flux ratios for compact sources
(angular sizes less than 4$\arcsec$) with peak fluxes
brighter than 0.2 mJy/beam and $R_{\rm min}\le 5\arcmin$ as a function of 
the radial distance from the center in
the final mosaic. 
 
\begin{figure}
\plotone{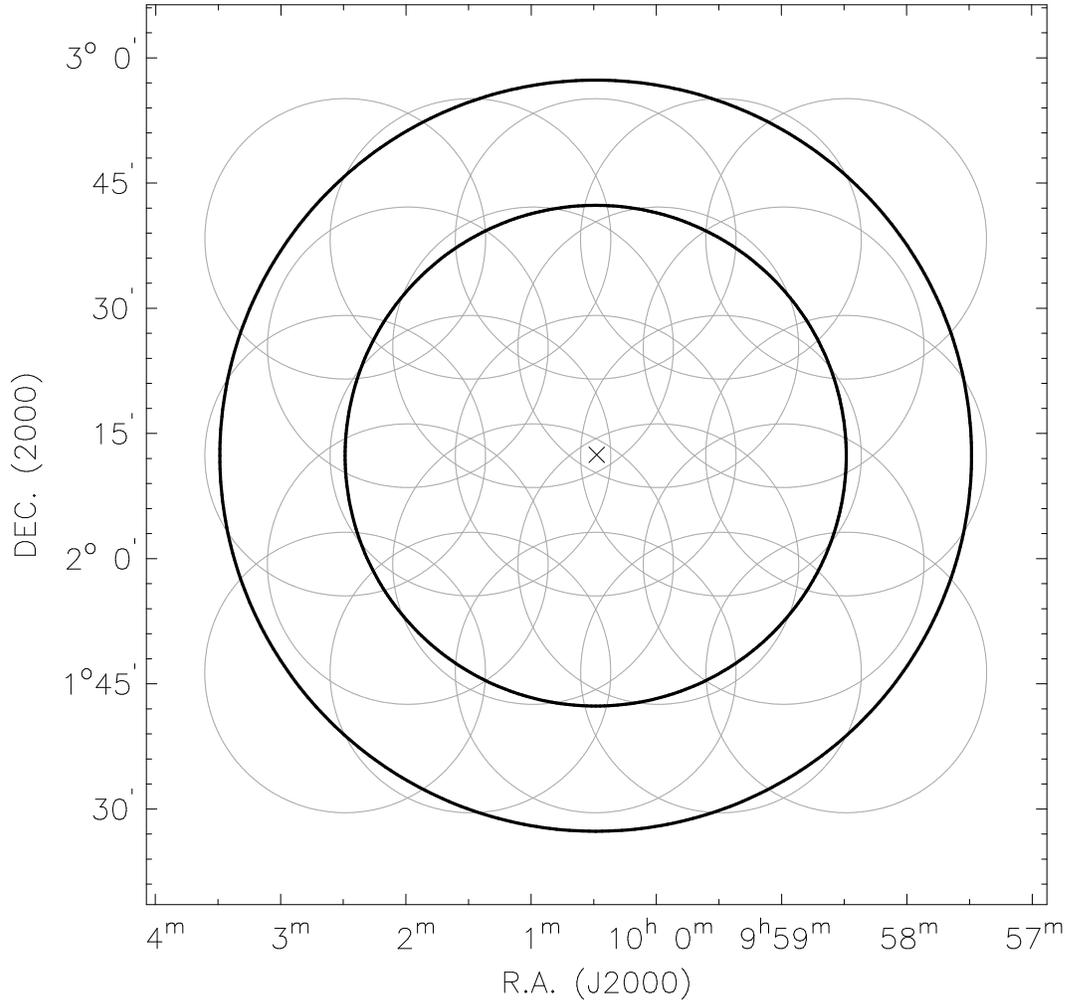}
\caption[]
{Layout of the 23 pointings for the VLA-COSMOS observations. The two circles
have a radius of $30\arcmin$ and $45\arcmin$. 
}
\label{pointings}
\end{figure}

\placefigure{pointings}

The ratio is about $0.8\pm 0.05$ for sources in the inner region of the 
mosaic, within 30$\arcmin$ from the mosaic center. 
In the region between 30$\arcmin$ and 45$\arcmin$, the ratio is $0.9\pm
0.03$, implying a smaller
correction in this area. This is due to the fact that the number of 
overlapping pointings is smaller at larger radii (
see Fig.~\ref{pointings}).
Within these two regions
the effect appears to be independent from the position in the mosaic considering the
dispersion of the points. Therefore, we adopted a constant bandwidth
smearing correction factor of 0.8 for the sources within $30\arcmin$ from the
mosaic center and a constant value of 0.9 for those between $30\arcmin$ and 
$45\arcmin$.

We performed this analysis in the inner 1 deg$^2$  ($1\times 1$ deg.
square region) of the final mosaic
since this region has the deepest and most uniform sensitivity and it
will be used to derive the source counts in the following Section.
Beyond $45\arcmin$ from the mosaic center this correction is no more valid as
each source in the mosaic receives a
contribution from only one pointing and therefore
the bandwidth smearing correction will
be a function of the distance from the center of the individual pointing
where each source is located, rather than  of the distance from the mosaic
center.

\begin{figure}
\plotone{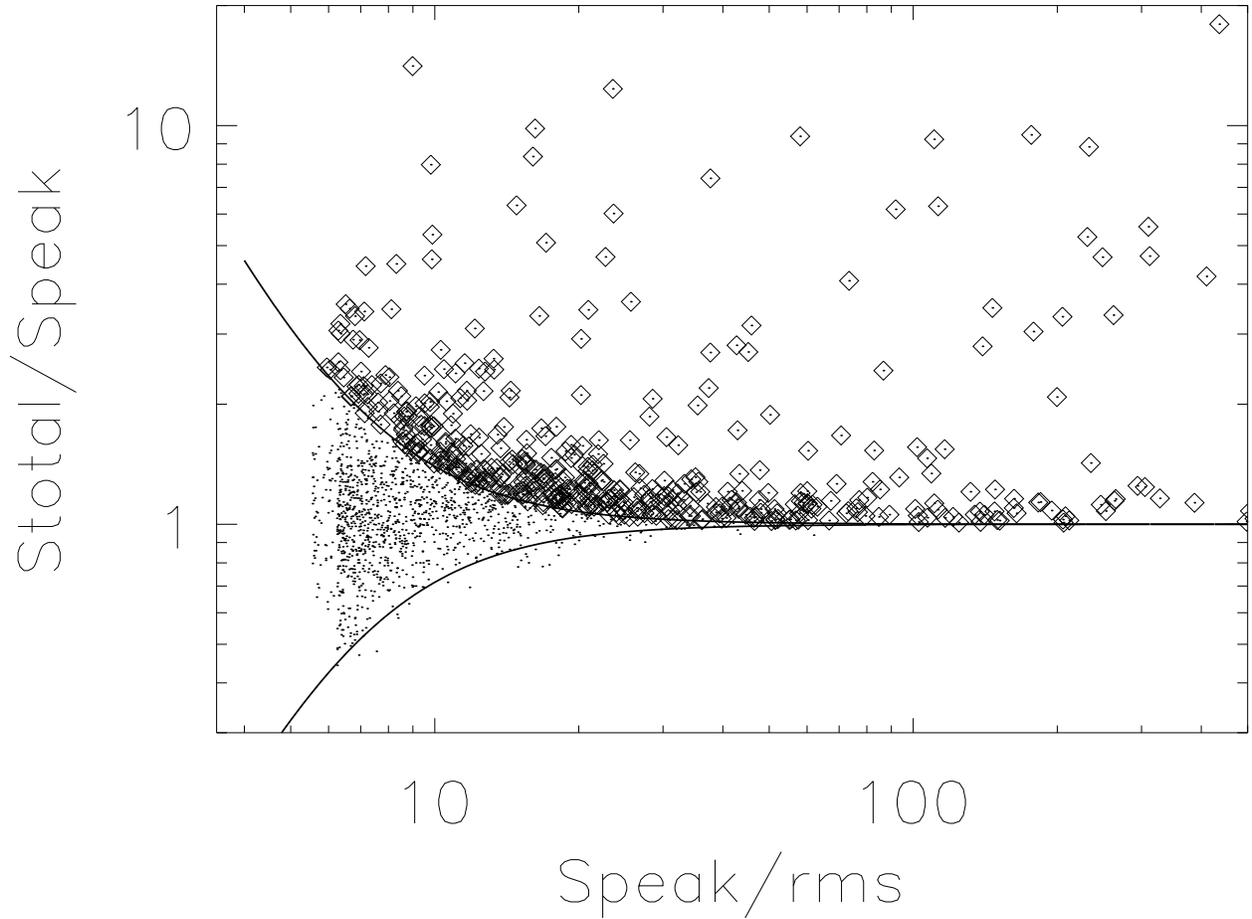}
\caption[]{Ratio of the total flux to the corrected peak flux (see text for
details) as a function of the
signal-to-noise ratio. The solid lines show the upper and lower envelopes of
the flux ratio distribution containing the sources considered unresolved.
Open symbols show the sources considered extended.
This Figure should be compared to Fig. 15 in S07. }
\label{res_unres}
\end{figure}

\placefigure{res_unres}

Consequently, to derive the radio source counts in the inner 1 deg$^2$
we divided all the 
peak flux densities by 0.8 or 0.9 depending on the distance from the
center of the mosaic and derived the new
resolved/unresolved criterion following the procedure described in S07. 
Since the ratio between total and peak fluxes is a direct measure of the
extent of a radio source, we used it to discriminate between resolved or
extended sources (i.e. larger than the beam) and unresolved sources.

Figure~\ref{res_unres} shows the 
ratio of the total integrated flux density $S_{\rm total}$ and the corrected
peak flux density $S_{\rm peak}$ as 
function of the signal to noise ratio. To select the resolved sources we have 
fitted a lower envelope in Fig.~\ref{res_unres} which contains $\simeq 95\%$ 
 of the sources with  $S_{\rm total} < S_{\rm peak}$ 
and mirrored it above the  $S_{\rm total}/ S_{\rm peak}=1$ line.
The criterion of 95\% of sources contained in the lower envelope of
Fig.~\ref{res_unres} is slightly different from that used in S07, where a
value of 99\% was adopted. We reckon that the value adopted here is the best
compromise to properly classify the sources as resolved or unresolved
because it fully characterises the shape of the distribution of sources with
$S_{\rm total} < S_{\rm peak}$ and, at the same time, it excludes the most 
extreme values. This envelope is described by the equation

\begin{equation}
S_{\rm total}/S_{\rm peak}= 1+ [100/(S_{\rm peak}/{\it rms})^{2.4}]
\end{equation}

It is important to note that in Fig.~\ref{res_unres} there is no systematic 
offset from the 
 $S_{\rm total}/ S_{\rm peak}=1$ line at high SNRs. This is  an a
 posteriori confirmation
 that we adopted an appropriate bandwith smearing correction.
Considering only sources with SNR$\ge 5$ in the inner 1 deg$^2$,
we obtain 484 resolved sources, for which the total flux is given by
the total flux of the Gaussian fitting, and 1208 
unresolved ones, for which the total flux is set equal to the peak flux.

It is worth  noting a few points regarding the effects of the correction 
on the VLA-COSMOS Large Project Catalog: 
\begin{itemize}
\item The correction has been determined only out to a radius of 45 arcmin
from the mosaic center and should be safely adopted only in this region.

\item
The catalog is still selected on the
old, pre-bandwidth smearing correction, peak flux densities or signal-to-noise
ratio, 
and we applied the bandwidth smearing correction only to the sources 
already in the catalog.
In the remaining we will refer to the signal-to-noise ratio as the 
original one, calculated with the uncorrected peak flux density.

\item
The correction affects the resolved/unresolved classification of a source
and therefore also the total flux density and size for the sources, previously 
classified as resolved, that are found unresolved after correcting for the
smearing. This affects about 500 sources, almost one third of the total
number of sources in the inner 1 deg$^2$ with 
SNR $\ge 5$.


\end{itemize}

\section{Survey completeness}

\subsection{The visibility area}
The VLA-COSMOS Large Project Catalog contains $\sim 3600$ radio sources 
extracted from a 2 deg$^2$
area. As shown in S07, the noise is not uniform over the whole area
and increases in the outer regions. The central 1 deg$^2$ area has
the lowest and most uniform noise (mean rms 10.5 $\mu$Jy) and  therefore
it is very well-suited
to derive the source counts. Extending the analysis to the whole area and 
catalog 
would  increase the number of objects at higher flux densities
($> 0.1-0.2$ mJy), but it would not add much to the accuracy of the source counts at
the lowest flux densities, which is the main focus of this paper. 
On the contrary, the correction factors that would need to be applied to take
into account the loss of the faintest sources in the outer and noiser region
would add more systematic uncertainties at the faint end of the source counts. 
Furthermore, in our following analysis we consider only
sources with signal to noise ratio greater than or equal to 5 (the VLA-COSMOS 
Large
Project Catalog has been selected with a signal to noise ratio threshold of
4.5) and without a flag for possible sidelobe residuals,
in order to minimize contamination from spurious sources.
Given these constraints, we have a catalog of 
1692 sources with SNR $\ge 5.0$ within the inner 1 deg$^2$ region.

\begin{figure}
\plotone{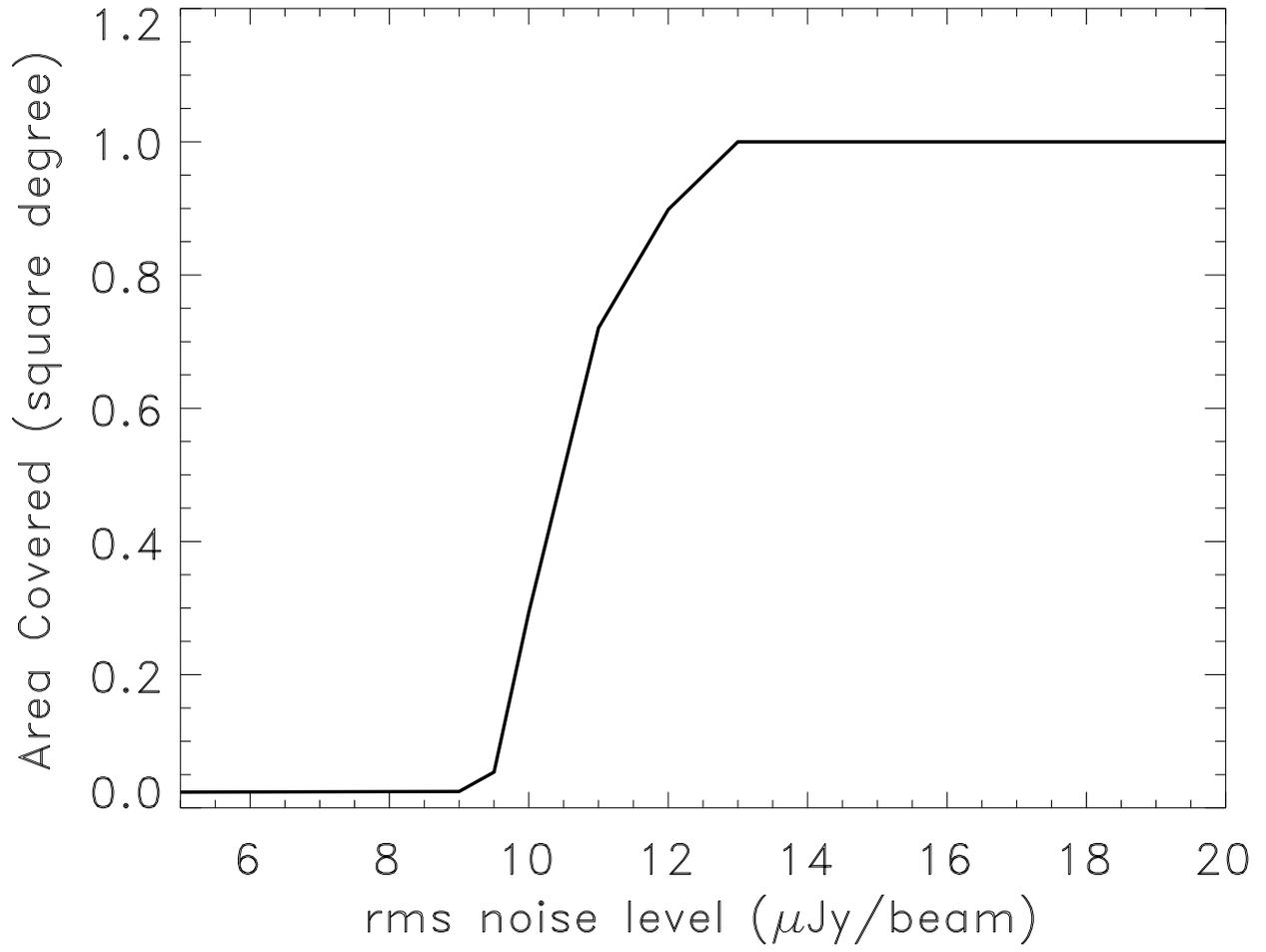}
\caption[]{Visibility area of the inner 1 deg$^2$ of the COSMOS survey.}
\label{visi_fig}
\end{figure}

\placefigure{visi_fig}

The visibility area of the inner 1 deg$^2$ of the VLA-COSMOS survey as a
function of the rms noise is shown in Fig.~\ref{visi_fig}.
This is the  area over which a source with given peak flux density
(5 times the noise in our case) can be detected.
Since the noise in this region is very smooth the visibility area increases
very rapidly and becomes flat above an r.m.s. noise of $\simeq 13$
$\mu$Jy/beam. The same curve, computed over the entire 2 deg$^2$, becomes
flat only above an r.m.s. noise of $\simeq 35$ $\mu$Jy/beam (see Fig.~13 in
S07).
This plot has been derived using the original peak flux
densities and not those corrected for the bandwidth smearing since the
catalog is still selected on the basis of the uncorrected peak flux.

\subsection{The resolution bias and the intrinsic angular size distribution
of sub-mJy radio sources}
Besides the visibility area of each source, other correction factors need to
be considered to estimate the completeness of the radio catalog and to
derive the intrinsic source counts: the
errors introduced by the fitting routines, the noise bias, and the resolution 
bias can all be modelled using simulated samples of radio sources. 
In particular,
the resolution bias can be quite important since the beam of the
VLA-COSMOS observations is  $1.5\arcsec \times 1.4\arcsec$ 
(e. g. about 15 times less 
in area than for similar surveys carried out at the same frequency).
Since the completeness of the radio catalog is defined in terms of the
peak flux, while the source counts are derived as a function of the
total flux density, corrections need to be applied to take into account
resolved sources with peak fluxes below the catalog threshold and
total flux densities above the nominal limit of the survey.

Our strategy to estimate the correction factor has been to simulate samples of
radio sources, with a realistic flux density and angular size distributions,
to insert these sources in the mosaic image and to recover these mock sources
using the same technique used to produce the catalog. From the comparison 
between the number of sources added to the image and those effectively
recovered we can estimate the correction factors for different flux density
bins.
The correct choice of the flux density distribution is rather
straightforward, as a broken power law derived from the observed 
distribution is a reasonable assumption, consistent with the extrapolation from
shallower surveys.
The choice of the angular size distribution for the simulated radio sources is
more uncertain.
While there is good evidence that the median of the intrinsic source size 
at 1 mJy is about $2\arcsec$ and decreases with
decreasing flux density at sub-mJy level, it is not known how exactly
this happens and what is
the best analytical relationship to model this behaviour.
As we will show later, this is an important parameter in deriving the
completeness of the VLA-COSMOS catalog.
In the past different scaling relationships between the median angular size,
$\theta_{\rm med}$, and the total flux density, $S$, have been used, e.g. 
$\theta_{\rm med}\propto S^m$ with $m=0.3$ \citep{Wind90} or 
$m=0.5$ \citep{Rich00}.

More recently, new surveys have provided some results on the 
angular size distributions of the $\mu$Jy radio sources.
\citet{Foma06} obtained a
complete sample of 289 radio sources with flux density $\ga$ 40 
$\mu$Jy and found that 
64\% of objects are  unresolved with sizes less than $1.2$ arcsec.  
MERLIN observations at higher angular resolution ($\sim 0.2\arcsec$) provided 
consistent results for
a sample of 92 radio sources, 80 of which with flux density
between 40 $\mu$Jy and 200 $\mu$Jy \citep{Muxl05}.
The angular size distribution derived from the latter sample of 
objects, which are all resolved, shows a dominant narrow gaussian component 
centered at $\simeq 0''$.7 and a tail
of sources ($\sim 20\%$ of the sample) with  angular sizes between 2$''$ 
and 4$''$.
We derived the median of the angular size distribution for the sources
observed by Muxlow et al. (2005) in the flux density range 40--90 $\mu$Jy.
The 55 sources in this flux density range have a median angular size of
$0\arcsec.7$ (with a r.m.s. dispersion of 0$\arcsec$.4) and a median flux
density of 54 $\mu$Jy. For comparison, \citet{Bond03} found that the
median angular size for sources in the flux density range
0.4-1.0 mJy is $1\arcsec.8$ with a median flux density of 0.56 mJy.
The value of $m$ consistent with these median angular sizes at different
flux densities is $m=0.4$. It is worth noting that this value for $m$ has 
to be considered as a lower limit
since the angular sizes given by \citet{Muxl05} are the largest angular sizes
while the angular sizes given in \citet{Bond03} are the FWHM.

These results provide strong evidences that radio sources with flux density
around 100 $\mu$Jy have typically sub-arcsecond sizes even if the quality of
the data (e.g. large number of upper limits in the angular sizes derived in 
Bondi et al. 2003) and the statistics (the above result on $m$ is derived from
less than 100 sources) do not allow one to derive a firm estimate of $m$.

Given these uncertainties, we decided to use a complementary method to 
derive the intrinsic angular size distribution, simulating mock samples of
radio sources with different angular size
distributions following a general power law $\theta\propto S^m$
with the exponent $m$ ranging from 0.2 to 0.5, and normalized to the integral
angular size distribution derived from the VVDS survey in \citet{Bond03}.

We constructed a simulated sample of radio sources containing about 3100
radio sources down to a total flux density 
level of 30 $\mu$Jy. This allows us to count also sources with flux density 
below
the limit which, because of positive noise fluctuations, might have a measured
peak flux density above the threshold. 
Then, all the simulated sources were randomly injected in the inner 1  
deg$^2$ of the VLA-COSMOS mosaic and subsequently recovered using the same
procedures adopted for the real sources and binned  in flux
density intervals. 
This was repeated for 3 different samples, and for four different 
angular size distributions (with $m$ equal to 0.2, 0.3, 0.4, and 0.5)
yielding a total of 12
simulated samples with more than 35,000 simulated radio sources.
Finally, from the comparison between the number of simulated sources
detected in each bin and the number of  sources in the simulated input
sample in the same flux density bin we derived the correction 
factors for each value of the exponent $m$. 
These correction factors, listed in Table~\ref{corr_5sigma},
account for the visibility area, noise bias, fitting errors
and resolution bias.

\placetable{corr_5sigma}

From Table~\ref{corr_5sigma} it is evident that
for the brightest sources the correction factors are $\la 10\%$ and 
the differences among  the correction factors for the various angular size 
distributions are negligible. On the other hand,
below $0.20$ mJy the correction factors become quite large with
significant differences for the various angular size distributions.
Therefore, it is important to infer the intrinsic angular size 
distribution in order to reconstruct the properly corrected source counts.
Another interesting feature of the correction factors which is worth noting 
is that the maximum is not at the lowest flux density bin but around 0.12 mJy.
This is due to a combination of the resolution bias and the capability to
classify a radio source as resolved, which depends on the signal-to-noise 
ratio. A source with a given total flux density and size will 
be catalogued in one of the following three classes:
1) undetected, if the peak flux is below the sample threshold,  
2) detected with the correct total flux density if the signal to noise ratio 
is high enough to allow for proper deconvolution, or, 
3) detected with a lower total flux density
if the signal to noise ratio is not high enough to recognize the source as
resolved. The sources in the last class are 
 assigned a total flux density lower than the intrinsic one and 
equal to the peak flux. A significant fraction of the sources with intrinsic 
total flux densities $\la 0.15$ mJy is redistributed to lower total flux
densities since the signal-to-noise ratio is not sufficient to allow a proper
deconvolution. 

\placetable{median_sim}

To constrain the intrinsic angular size distribution we performed the
following test. 
We derived the median major axis of the simulated samples
of radio sources for each of the tested power laws ($m=0.2-0.5$) in the
flux density range 0.25-0.4 mJy. 
This flux density range was chosen as a compromise to minimise the
number of unresolved sources (and therefore upper limits on source size) 
and allow for a 
statistically significant large sample of sources reaching as close as 
possible to the lowest fluxes.
We calculated both the median major axis  of the original
injected sources (row 1 in Tab.~\ref{median_sim}) and the median major axis
of the simulated sources recovered by the fitting routine (row 2 in
Tab.~\ref{median_sim}). These latter values were obtained considering also the
upper limits derived for the sources fitted as unresolved 
($\la$ 15\% of the total number of sources in this flux 
density range). 

Table~\ref{median_sim} shows that,
in the flux density range 0.25-0.4 mJy, 
although the measured median major axis of the simulated sources is
somewhat larger than that actually injected, the differences are not
statistically significant for any given 
value of $m$. 
On the other hand, different values of $m$ yield to
significantly different median major axis.

Therefore, the value of the median major axis in this flux density range for 
the real sources could be used to roughly discriminate between different 
intrinsic angular size distributions. Unfortunately, the measured sizes
of the radio sources are affected by bandwidth smearing and they can not be
used for such a test. However, as we have seen in Section 2, the
total flux densities are properly determined and the ratio between the
total and the corrected peak flux densities can be used as an estimate of the
area covered by the radio sources. Then we made the assumption that
radio sources are extended in only one direction and we derived the major axis
from the ratio between the total and the corrected peak flux densities.
This assumption can be justified since extended radio emission is 
generally associated to elongated features (e.g. radio jets in AGN or spiral 
arms in star-forming galaxies), and when this is not true the derived source
sizes will be an upper limit of the real ones.
In this way we obtained a median value of the major axis
for the observed sources in the flux density range 0.25-0.4 mJy of 
$0\arcsec.99\pm 0\arcsec.09$.
From the comparison with the values listed in Tab.~\ref{median_sim}, we 
conclude that the observed median major axis 
is more consistent with that obtained using simulated samples of 
radio sources following 
an angular size distribution with $m=0.5$ or $m=0.4$.

With both  methods, the comparison between the angular size distribution
obtained by \citet{Muxl05} and \citet{Bond03} in different flux density
intervals and the comparison between the estimate of the true angular sizes
and those derived from the simulated samples, we obtained consistent results
which exclude values of $m=0.2$ and $m=0.3$ and suggest $m\simeq 0.5$. In the 
following analysis we apply the correction factors obtained for $m=0.5$  
to the radio source counts. 

\section{The radio source counts}

We have derived the radio source counts at 1.4 GHz in the inner 1 deg$^2$
region of the COSMOS field
from the VLA-COSMOS catalog, corrected for the effect of bandwidth smearing 
and with a signal-to-noise threshold of 5.
The source counts up to $\simeq 1$ mJy are presented in 
Table~\ref{counts_tab}. For each flux
density bin we give the mean flux density, the observed
number of sources, the corresponding differential source
density d$N$/d$S$ (in sr$^{-1}$Jy$^{-1}$), the normalised differential counts 
(d$N$/d$S$)$S^{2.5}$ (in  sr$^{-1}$Jy$^{1.5}$) with their estimated Poissonian 
errors, the incompleteness correction factors from
Table~\ref{corr_5sigma} for $m=0.5$, and the cumulative number of effective
sources after applying the corrections. The correction factors above 
0.2 mJy are $\le 10\%$. Such variations are roughly consistent with
the poissonian error of the source counts and for this reason the source
counts above 0.2 mJy are assigned a correction factor of 1.
The mean flux density in each bin was calculated as the 
geometric mean of the flux density extrema.

\placetable{counts_tab}

\begin{figure}
\plotone{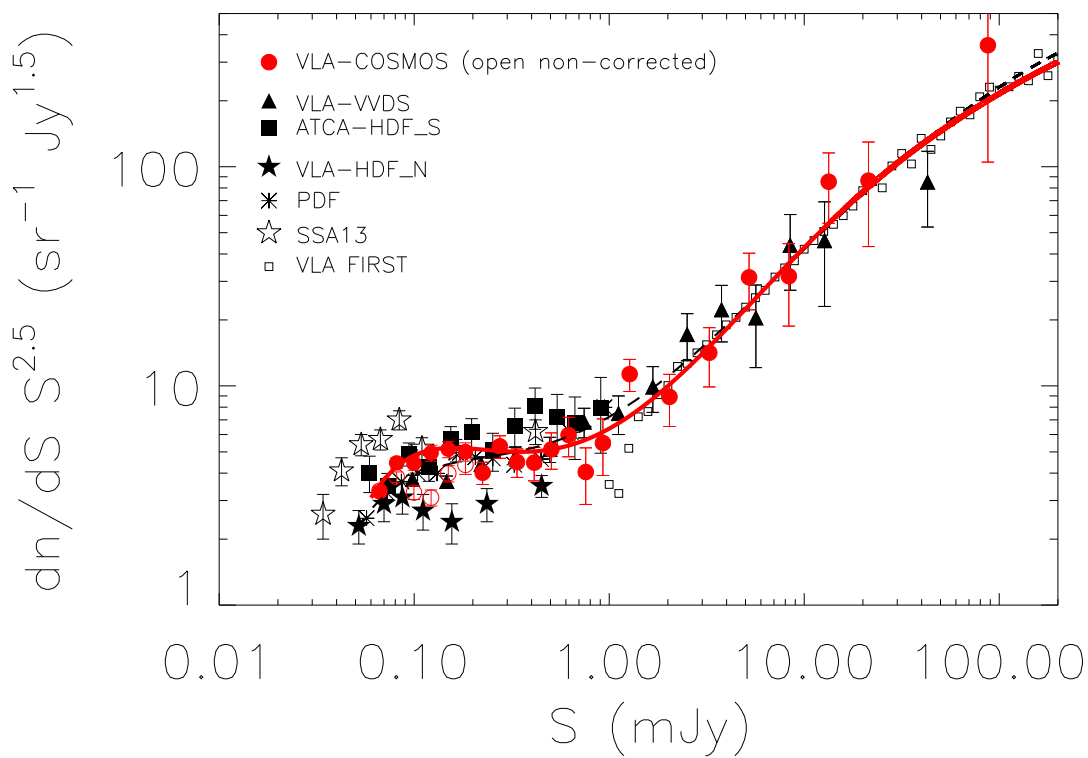}
\caption[]{Radio source counts at 1.4 GHz from the VLA-COSMOS survey (dots)
and from other surveys. Empty circles show the radio counts not corrected for
incompleteness, filled circles the corrected ones using $m=0.5$. 
The VLA-COSMOS source counts
are shown along with those obtained by other deep surveys (see text). The
solid line is least-squares sixth-order polynomial fit obtained using the
VLA-COSMOS and the FIRST source counts. The dashed line is the fit obtained
by \citet{Hopk03}.}
\label{counts1}
\end{figure}

\begin{figure}
\plotone{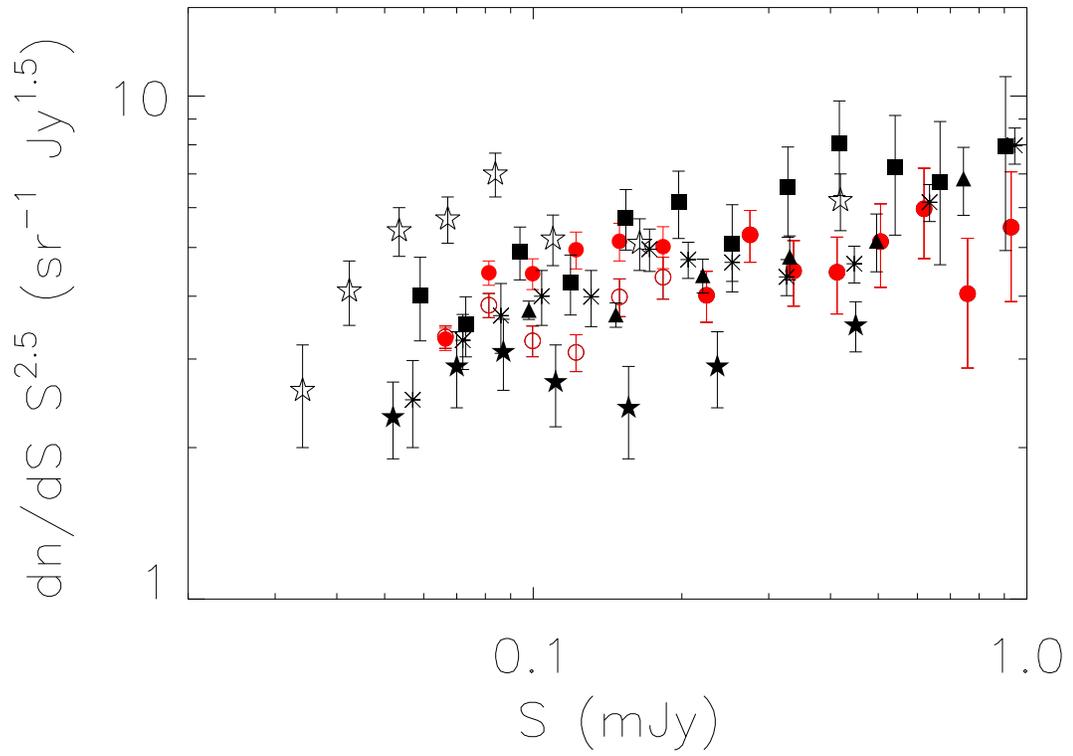}
\caption[]{Blow-up of the radio source counts in the sub-mJy region.
Symbols are the same as in Fig.~\ref{counts1}.}
\label{counts2}
\end{figure}

In Fig.~\ref{counts1} and Fig.~\ref{counts2} we show
the raw and the corrected  source counts derived from 
the VLA-COSMOS survey compared to those from other surveys at 1.4 GHz:
the VLA-VVDS \citep{Bond03}, the ATCA-HDF\_S
\citep{Huyn05}, the VLA-HDF\_N \citep{Rich00}, the PDF survey
\citep{Hopk03}, the SSA 13 field \citep{Foma06} and the FIRST survey
\citep{Whit97}.
The full range of source counts up to $\simeq 100$ mJy is shown in
Fig.~~\ref{counts1}, and a blow-up of the sub-mJy region is shown in
Fig.\ref{counts2}. In both figures the raw source counts from the VLA-COSMOS
survey are shown with empty circles and those corrected for incompleteness
with filled ones.
The solid line in Fig.~\ref{counts1} is a linear least-squares
sixth-order polynomial fit obtained using the VLA-COSMOS source counts
above 0.06 mJy supplemented with the FIRST source counts above 2.5 mJy and
up to 1 Jy. The resulting polynomial fit is given by

\begin{equation}
\log[(dN/dS)/(S^{-2.5})]=\sum_{i=0}^6 a_i[\log(S/mJy)]^i
\end{equation}

with $a_0=0.805$, $a_1=0.493$, $a_2=0.564$, $a_3=-0.129$, $a_4=-0.195$,
$a_5=0.110$, and $a_6=-0.017$.
The residuals from the polynomial fit have an rms of about 0.06 in the
logarithm of the normalized counts.
As discussed by \citet{Hopk03} the sixth-order
polynomial is necessary to follow the different curvatures shown by the source
counts.
In Fig.~\ref{counts1} we also show for comparison with a dashed line 
the sixth-order polynomial fit obtained by \citet{Hopk03} from the Phoenix
Deep Survey source counts. The two fits are in reasonably good agreement with
our fit slightly higher for flux density below 0.2 mJy.

\placefigure{counts1}
\placefigure{counts2}

It is evident that in some cases, the field-to-field differences in the radio 
source counts below 0.5 mJy are larger than the combined errors. 
It is likely
that these differences (at least partly) can be ascribed 
to instrumental defects or different recipes to account for the incompleteness.
In particular, the assumption of the intrinsic angular size
distribution of the $\mu$Jy radio sources in high resolution radio surveys is
certainly a factor, as we have shown in the previous Section. Nevertheless,
the most extreme cases (e.g. the SSA 13 and HDF North fields) should reflect
real cosmic variance \citep{Foma06}.

There is a particularly good agreement between the source counts derived
from the VLA-COSMOS, after correction, and the HDF-S fields. This is 
reassuring for the
resolution bias correction we applied since the HDF-S field was observed with
a resolution of about $5\arcsec$, and for this reason 
is  much less affected by the precise
shape of the intrinsic angular size distribution.

It is worth noting that while the VLA-COSMOS survey is not as deep as 
the SSA13 and HDF-N surveys, it is, by far, the one with the most robust
statistics. At a flux density level of $\sim 65$ $\mu$Jy the VLA-COSMOS
survey counts almost 400 sources while typical numbers for other surveys,
in this flux density range, are  between 50 and 90 sources.  

A possible drop off in the radio source counts below $\sim 100$ $\mu$Jy has 
been already detected and discussed by other surveys 
imaging various fields with different sensitivity and resolution
\citep[e.g.]{Hopk03, Huyn05, Foma06}.
Although, in every case, the drop off relies on the faintest bins it should 
not be immediately  discounted as incompleteness in the source counts as 
it is accepted that the flat region of the normalised source counts below
$\sim 0.3$ mJy can not continue indefinetely \citep{Wind93}.
From the VLA-COSMOS radio source counts there is some indication that the 
density  of sources is beginning to decrease below 100-150 $\mu$Jy producing 
a downturn in the observed radio source counts, even if the result is 
strongly dependent on the fidelity of the first flux density bin.
It is important to note that if incompleteness is responsible for the drop
off observed in the first flux density bin of Fig.~\ref{counts2}, it would
mean that we are missing about 130 sources in the range 0.06-0.0735 mJy, which
is extremely unlikely. So, while incompleteness at the lowest flux
density levels can still be present at some level, it can not explain by 
itself the drop off observed in the radio source counts. Thus, this decline
can provide important limits when modeling the
populations contributing to the radio source counts.

As we have explained in the previous section, the intrinsic angular size
distribution adopted to derive the incompleteness correction factor has a
large impact in this flux density range, but adopting distributions with
$m< 0.5$ would only enhance the drop off below $\simeq 0.15$ mJy.

Also the uncertainties in the bandwidth smearing correction could, in
principle,  significantly
modify the radio source counts and the observed drop off observed at lower 
flux density. We have tested
this assuming two extreme cases, a constant correction over the whole
area of 0.8 and 0.9 respectively. As we can expect,
the first case would produce a very small
difference, well within the Poissonian error boxes, since we already applied 
the 0.8 correction over almost 80\% of the 1 deg$^2$. On the other hand, 
adopting a correction
factor of 0.9 over the whole area, would significantly lower only the radio 
source counts in the first three bins, increasing the slope of the drop off 
region. As we have shown in Section 2, the assumption of a constant correction
over the whole 1 deg$^2$ is not corroborated by the real data. The data
require a mixture of the two correction factors and with the previous test we 
have just verified that the drop off in the source counts below 0.15 mJy can 
not be produced by the applied smearing corrections.

We can conclude that given the large statistics and the fact that the 
incompleteness factors
have been carefully investigated and applied, the decrease in the
source counts below $\simeq 0.10-0.15$ mJy observed in the VLA-COSMOS radio
source counts should be considered real.

\section{Summary \& conclusions}

In this paper we have presented a further analysis of the VLA-COSMOS Large
Project Catalog
with the goals of: 1) quantifying and correcting for the bandwidth smearing 
effect which affects the VLA-COSMOS radio mosaic and catalog; 
2) determining the incompleteness of the VLA-COSMOS
radio catalog, through extensive use of Monte-Carlo simulations 
taking into account different intrinsic angular size distributions, 
in order to
quantify the effects of noise bias and resolution bias; 3) deriving the radio
source counts at 1.4 GHz in the VLA-COSMOS field.

The bandwidth smearing correction factors have been determined from the
comparison of peak and total flux densities in the final mosaic and
in individual pointings where the source is within $5\arcmin$ from the center.
We have found that a two-value correction factor is a good approximation: 
out to a radius of $30\arcmin$ from the center,
the peak flux densities in the VLA-COSMOS catalog must be 
divided by a factor 0.8, while in the region between $30\arcmin$ and 
$45\arcmin$ from the
center by a factor 0.9. Inside these two regions the correction
factor is rather constant, within the dispersion of the data, with a sharp
transition from 0.8 to 0.9 at a radial distance of $30\arcmin$ reflecting
the change in overlapping pointings.
Consequently, a new classification in resolved and unresolved sources has been
derived. A number of sources,
originally classified as resolved, turned out to be unresolved 
after the bandwidth smearing correction was applied. For these sources
also the total flux density was corrected.

We simulated several samples of radio sources with a flux density
distribution compatible with the observed one and with different 
angular size distributions. 
The angular size distribution was modelled using $\theta\propto S^m$
with $m=0.2,0.3,0.4,0.5$.
These mock sources were added to the radio mosaic 
and recoverd using the same procedure that yielded to the radio catalog.
In such a way we were able to quantify the effects of the combined noise and
resolution bias affecting the completeness of the radio catalog.
A comparison of the median angular size at a flux density level
not significantly affected by the resolution bias for the simulated samples of
sources and the observed one, strongly suggest a general distribution of
radio source sizes following $\theta\propto S^{0.5}$.

Finally, we derived the radio source counts in the VLA-COSMOS field (both
uncorrected and corrected for the incompleteness). The source counts extend
down to
60 $\mu$Jy and are consistent with those derived by other surveys but with 
much more robust statistics. In particular, the drop off of the source counts
below $\sim 100$ $\mu$Jy has to be considered real and not due to
incompleteness.

\begin{acknowledgements}
The National Radio Astronomy Observatory (NRAO) is operated by 
Associated Universities, Inc., under cooperative agreement with the 
National Science Foundation. KJ acknowledges support by the German DFG 
under grant SCHI~536/3-1.
\end{acknowledgements}

\begin{table}
\centering
\caption{Correction factors for different angular size distributions}
\label{corr_5sigma}
\begin{tabular}{cccccccc}
\hline
Flux bin        &  $m=0.2$  &  $m=0.3$  &  $m=0.4$  & $m=0.5$ \\
(mJy)           &           &           &           &          \\
\hline
0.0600-0.0735   &   1.29    &   1.19    &  1.06     & 0.99  \\     
0.0735-0.0900   &   1.59    &   1.35    &  1.23     & 1.16  \\     
0.0900-0.1103   &   2.12    &   1.81    &  1.53     & 1.36  \\
0.1103-0.1351   &   2.34    &   2.12    &  1.81     & 1.60  \\
0.1351-0.1655   &   1.60    &   1.46    &  1.32     & 1.29  \\
0.1655-0.2028   &   1.24    &   1.21    &  1.21     & 1.15  \\
0.2028-0.2484   &   1.05    &   1.03    &  1.00     & 0.99  \\
0.2484-0.3043   &   1.03    &   1.02    &  0.95     & 0.93  \\
0.3043-0.4564   &   1.10    &   1.08    &  1.09     & 1.07  \\
0.4564-0.6846   &   1.11    &   1.10    &  1.10     & 1.09  \\
0.6846-1.0270   &   0.98    &   0.98    &  0.97     & 0.98  \\
$\ge 1.0270$    &   1.07    &   1.07    &  1.08     & 1.09  \\
\hline
\end{tabular}
\end{table}

\begin{table}
\centering
\caption{Median major axis in the flux density range 0.25-0.4 mJy for the
input and output simulated samples}
\label{median_sim}
\begin{tabular}{cccccccc}
\hline
   Sample       &  $m=0.2$    &  $m=0.3$  &  $m=0.4$  & $m=0.5$ \\
   &$\theta_{maj,med}$&$\theta_{maj,med}$&$\theta_{maj,med}$&$\theta_{maj,med}$\\
\hline
Sim. Input      &$1.32\pm 0.05$&$1.17\pm 0.04$&$1.03\pm 0.04$&$0.91\pm 0.04$\\
Sim. Output     &$1.36\pm 0.06$&$1.24\pm 0.06$&$1.11\pm 0.05$&$1.00\pm 0.05$\\
\hline
\end{tabular}
\end{table}

\begin{table}
\centering
\caption{The 1.4 GHz radio source counts}
\label{counts_tab}
\begin{tabular}{cccccccc}
\hline
 $S$ &  $<S>$ & $N$& d$N$/d$S$& (d$N$/d$S$)$S^{2.5}$ & C & N$_{c}$($>S$)\\
 mJy &  mJy   &     &sr$^{-1}$Jy$^{-1}$& sr$^{-1}$Jy$^{1.5}$& & deg$^{-2}$\\
\hline
0.0600-0.0735   & 0.066 & 380  & $9.24\times 10^{10}$& $3.32\pm 0.17$ & 0.99
& $1669\pm 41$\\
0.0735-0.0900   & 0.081 & 324  & $6.43\times 10^{10}$& $3.84\pm 0.21$ & 1.16
& $1515\pm 39$\\
0.0900-0.1103   & 0.100 & 203  & $3.29\times 10^{10}$& $3.26\pm 0.23$ & 1.36
& $1336\pm 37$\\
0.1103-0.1351   & 0.122 & 142  & $1.88\times 10^{10}$& $3.09\pm 0.26$ & 1.60
& $1246\pm 35$\\
0.1351-0.1655   & 0.150 & 135  & $1.46\times 10^{10}$& $3.99\pm 0.34$ & 1.29
& $~822\pm 29$\\
0.1655-0.2028   & 0.183 & 109  & $9.61\times 10^{9}$ & $4.36\pm 0.42$ & 1.15
& $~577\pm 24$\\
0.2028-0.2484   & 0.224 & ~74  & $5.33\times 10^{9}$ & $4.02\pm 0.47$ & 1.00
& $~393\pm 20$\\
0.2484-0.3043   & 0.275 & ~72  & $4.23\times 10^{9}$ & $5.30\pm 0.63$ & 1.00
& $~319\pm 18$\\
0.3043-0.3728   & 0.337 & ~45  & $2.16\times 10^{9}$ & $4.49\pm 0.67$ & 1.00
& $~247\pm 16$\\
0.3728-0.4566   & 0.413 & ~33  & $1.29\times 10^{9}$ & $4.47\pm 0.78$ & 1.00
& $~202\pm 14$\\
0.4566-0.5593   & 0.505 & ~28  & $8.95\times 10^{8}$ & $5.14\pm 0.97$ & 1.00
& $~169\pm 13$\\
0.5593-0.6851   & 0.619 & ~24  & $6.26\times 10^{8}$ & $5.97\pm 1.28$ & 1.00
& $~141\pm 12$\\
0.6851-0.8393   & 0.758 & ~12  & $2.56\times 10^{8}$ & $4.05\pm 1.17$ & 1.00
& $~117\pm 11$\\
0.8393-1.0282   & 0.929 & ~12  & $2.09\times 10^{8}$ & $5.49\pm 1.58$ & 1.00
& $~105\pm 10$\\
\hline
\end{tabular}
\end{table}

\clearpage

\end{document}